\documentclass[twocolumn,pra,aps,amssymb,amsfonts]{revtex4}
\usepackage{times,graphicx}
\newcommand{\trace}{\mathop{\rm Tr}\nolimits}

\newcommand{\qed}{\hfill$\square$\par\vskip24pt}

\newcommand{\cH}{{\cal H}}

\newcommand{\C}{{\mathbb{C}}}

\newcommand{\id}{\mathrm{\openone}}
\newcommand{\twomat}[4]{\left(\begin{array}{cc}#1&#2\\#3&#4\end{array}\right)}

\newcommand{\be}{\begin{equation}}
\newcommand{\ee}{\end{equation}}
\newcommand{\bea}{\begin{eqnarray}}
\newcommand{\eea}{\end{eqnarray}}
\newcommand{\beas}{\begin{eqnarray*}}
\newcommand{\eeas}{\end{eqnarray*}}

\begin{document}
\title{A Note on the $p\rightarrow q$ norms of Completely Positive Maps}
\author{Koenraad M.R. Audenaert}
\email{k.audenaert@imperial.ac.uk}
\affiliation{Blackett Laboratory, Imperial College London,
Prince Consort Road, London SW7 2BW, United Kingdom}
\affiliation{Institute for Mathematical Sciences, Imperial College London,
Exhibition Road, London SW7 2BW, United Kingdom}
\date{\today}
\begin{abstract}
King and Ruskai asked
whether the $p\rightarrow q$ norm of a completely positive map $\Phi$,
acting between Schatten $p$ and $q$ classes of self-adjoint operators,
$$
||\Phi||_{p\rightarrow q} = \max_{A=A^*} \frac{||\Phi(A)||_q}{||A||_p},
$$
is equal to the $p\rightarrow q$ norm of that map when acting between Schatten classes of general, not necessarily
self-adjoint, operators.
The first proof has been given by Watrous.
We give an alternative proof of this statement.
\end{abstract}
\maketitle
The Schatten $q$-norm is the non-commutative generalisation of the $\ell_q$ norms: for an operator $A$
and for $q\ge 1$
$$
||A||_q := \left(\trace(|A|^q)\right)^{1/q},
$$
where the absolute value $|A|$ is defined as
$$
|A| = (A^* A)^{1/2}.
$$

The maximal output purity of a completely positive linear map $\Phi$, as measured by the Schatten $q$-norm, is defined as
$$
\nu_p(\Phi) = \max_{A>0\atop \trace(A)=1} ||\Phi(A)||_q.
$$
This quantity is a measure of ``noisiness'' of the map \cite{ah}, considering it as a channel for the quantum state $A$.
It expresses the fact that a pure state will acquire at least this amount of mixedness when passing through the channel.

It has been proven in \cite{ah} that this quantity is equal to the $1\rightarrow q$ norm of $\Phi$,
considering it as a map from the Schatten 1-class of self-adjoint operators to the Schatten $q$-class.
That is
$$
\nu_p(\Phi) = \max_{A=A^*}\frac{||\Phi(A)||_q}{||A||_1},
$$
where the condition of positivity of $A$ is no longer required.

More generally, one can define the $p\rightarrow q$ norm
as
$$
||\Phi||_{p\rightarrow q} := \max_{A=A^*} \frac{||\Phi(A)||_q}{||A||_p}.
$$
King and Ruskai asked in \cite{kr} whether the condition of self-adjointness of $A$ in this definition
can also be dropped, that is, whether the following holds:
$$
||\Phi||_{p\rightarrow q} = \max_{A} \frac{||\Phi(A)||_q}{||A||_p}.
$$
They have proved this in \cite{kr} for the case $p=q=2$, and for the case of completely positive trace preserving maps
between 2-dimensional Schatten classes with $p=1$ and $q\ge 2$.

A completely general proof has been given by Watrous \cite{wat} for any completely positive map and
for all values of $p,q\ge 1$. In this note we present an alternative proof.

\textit{Proof.}
Our proof uses the construction \cite{bhatia} of turning a general operator $A$ on a Hilbert space $\cH$
into a self-adjoint one by defining an operator on the doubled space $\cH\oplus\cH$
$$
Q := \twomat{0}{A}{A^*}{0},
$$
which is obviously self-adjoint.

The absolute value of $A$ is defined as $|A|=(A^* A)^{1/2}$, and there exists a unitary $U$ such that
$(AA^*)^{1/2} = U|A|U^*$. The absolute value of $Q$ is therefore
\beas
|Q| &=& \left[\twomat{0}{A}{A^*}{0}\,\,\twomat{0}{A}{A^*}{0}\right]^{1/2} \\
&=& \twomat{AA^*}{0}{0}{A^*A}^{1/2} \\
&=& \twomat{U|A|U^*}{0}{0}{|A|}.
\eeas

The doubled space $\cH\oplus\cH$ is isomorphic to the space $\cH\otimes \C^2$.
Let $\Phi\otimes\id_2$ be the completely positive map extending $\Phi$ that operates trivially on $\C^2$.
Then, using the fact that for a completely positive map $\Phi(A^*) = \Phi(A)^*$,
$$
(\Phi\otimes\id_2)(Q) = \twomat{0}{\Phi(A)}{\Phi(A)^*}{0}.
$$
Thus
\beas
\lefteqn{||(\Phi\otimes\id_2)(Q)||^q_q} \\
&=& \trace|(\Phi\otimes\id_2)(Q)|^q \\
&=& \trace\left|\twomat{0}{\Phi(A)}{\Phi(A)^*}{0}\right|^q \\
&=& \trace\twomat{(\Phi(A)\Phi(A)^*)^{1/2}}{0}{0}{(\Phi(A)^*\Phi(A))^{1/2}}^q \\
&=& 2\trace|\Phi(A)|^p,
\eeas
where in the last line we have again used unitary equivalence of $XX^*$ and $X^*X$.

Likewise,
\beas
||(\Phi\otimes\id_2)(|Q|)||^q_q &=& \trace(\Phi\otimes\id_2)(|Q|)^q \\
&=& \trace\twomat{\Phi(U|A|U^*)}{0}{0}{\Phi(|A|)}^q \\
&=& \trace\Phi(U|A|U^*)^q+\trace\Phi(|A|)^q.
\eeas
Here we have used complete positivity of $\Phi$, implying positivity of $(\Phi\otimes\id_2)(|Q|)$.

Amosov and Holevo have shown in \cite{ah} that,
for any completely positive map $\Omega$, and any self-adjoint operator $X$,
$||\Omega(X)||_q \le ||\Omega(|X|)||_q$. Applying this to the map $\Phi\otimes\id_2$ and the operator $Q$,
and combining with the above results gives
$$
2\trace|\Phi(A)|^q \le \trace\Phi(U|A|U^*)^q+\trace\Phi(|A|)^q.
$$

Let us now impose the requirement $||A||_p=1$.
Thus, obviously, both $|A|$ and $U|A|U^*$ have $p$-norm equal to 1.
Then, by definition of the $p\rightarrow q$ norm of $\Phi$,
\beas
\trace\Phi(U|A|U^*)^q &\le& ||\Phi||_{p\rightarrow q}^q \\
\trace\Phi(|A|)^q &\le& ||\Phi||_{p\rightarrow q}^q,
\eeas
so that also
$$
2\trace|\Phi(A)|^q \le 2||\Phi||_{p\rightarrow q}^q.
$$
For general $A$, $A/||A||_p$ has $p$-norm equal to 1, so that by homogeneity,
$$
\frac{||\Phi(A)||_q}{||A||_p} \le ||\Phi||_{p\rightarrow q}.
$$
Maximising over all operators $A$, which includes the self-adjoint ones, we find
$$
\max_{A}\frac{||\Phi(A)||_q}{||A||_p} = ||\Phi||_{p\rightarrow q}.
$$
\qed


\end{document}